# Access Service Records Based Trust Management Scheme for Internet of Things

Xuefei Li, *Member, IEEE*, Ru Li, *Member, IEEE*

*Abstract*—**The distributed structure of the Internet of things has gradually replaced the centralized structure because of its scalability, security and single point of failure. The huge scale of information recording of the Internet of things brings challenges and opportunities to the trust management of the Internet of things. Through the analysis of a variety of existing trust management schemes, this paper proposes a unified data structure for distributed access service records, TokenChain, in which triple DES data encryption is used to ensure privacy and traceability, in order to achieve a unified trust management scheme. Based on TokenChain, a three-tier trust management architecture (TokenChain-Based Trust Management, TBTM) is implemented in the data layer, computing layer and control layer. Trust evaluation is affected by four statistics and finally converges to a real value under certain conditions. Based on TBTM, we carry out theoretical analysis, malicious attack resistance, simulation and performance evaluation in a variety of complex scenarios. The results show that TBTM satisfies service prediction, global trust analysis, high security and excellent performance in multi-domain complex scenarios. Compared with the existing trust management schemes, TBTM realizes cross-scenario interaction and two-way trust management, and simultaneously meets the characteristics of traceability, tamper-proof, attack resistance and distribution. Finally, this paper is summarized.**

## I. INTRODUCTION

The Internet of things is an emerging technology of the network physical system, which is a huge network composed of a large number of sensors, servers, clients, mobile devices and so on. It not only brings opportunities and convenience to life, but also brings a large number of threats and challenges. The distributed Internet of things has replaced the centralized Internet of things because of its superior distributed characteristics and expansibility [3]. The deep coupling between the Internet of things and the blockchain [6] can be used in intelligent manufacturing, supply chain management, vehicle networking, UAV and so on. But at the same time, it is also accompanied by security issues such as resource constraints, privacy disclosure and so on. Novo et al. [1] implemented a distributed Internet of things access control by issuing intelligent contracts to the blockchain, which shows that the blockchain has broad prospects and advantages in the application of Internet of things access control. The Internet of things is gradually integrated into our lives, enriching life through intelligent interactions between devices and man-machine, on the basis of which the social Internet of things and industrial Internet of things are derived. Vlacheas et al. [5] described a case of a smart city scenario. By analyzing the information exchange and self-reconstruction between composite virtual objects (including sensors, local area network platforms, etc.), it reveals the great potential of the reconfigurable Internet of things in the development of smart cities.

Blockchain [10] is a typically distributed ledger technology, which ensures the consistency of data through the consensus mechanism among multiple nodes [11], and ensures that the data can not be usurped (only added, not deleted and modified) through Merkle root, and the data of blockchain is traceable. Many researchers have great interest in blockchain [16] [18], especially in the study of consensus mechanisms. Kaur et al. [12] proposed an evaluation matrix of consistency protocol, which helps researchers to select appropriate consensus mechanism through different performance parameters. at present, there are about 130 kinds of distributed consensus mechanisms [13]. At the same time, there are endless studies on the combination of blockchain and trust management [2][8], the combination of blockchain and access control [1][14], the combination of blockchain and vehicle networks [15], and so on. The intelligent contract is a kind of contract code running on the blockchain platform, which has the advantages of automatically executing and processing transactions on the chain. YuanyuZhang et al. [14] proposed an access control scheme based on intelligent contracts. Simulation experiments show that intelligent contract can be used to achieve resource access control, and has the advantage of automation. Users can

This paragraph of the first footnote will contain the date on which you submitted your paper for review, which is populated by IEEE. This paper is supported by the National Natural Science Foundation of China (61862046).The project （2021-KJXM-TZJW-04 ）is supported by Huhhot Science & Technology Plan.This paper is supported by the Science and Technology Program of Inner Mongolia Autonomous Region (2020GG0188).

Xuefei Li is with school of Computing, Inner Mongolia University, Hohhot 010021, China (e-mail: 32109035@mail.imu.edu.cn; 15049163676@163.com)

Ru Li is with school of Computing, Inner Mongolia University, Hohhot 010021, China (e-mail: csliru@imu.edu.cn; 715783148@qq.com.)

Second B. Author, Jr., was with Rice University, Houston, TX 77005 USA. He is now with the Department of Physics, Colorado State University, Fort Collins, CO 80523 USA (e-mail: lamar@lamar.colostate.edu).

Third C. Author is with the Electrical Engineering Department, University of Colorado, Boulder, CO 80309 USA, on leave from the National Research Institute for Metals, Tsukuba 305-0047, Japan (e-mail: author@nrim.go.jp).

Mentions of supplemental materials and animal/human rights statements can be included here.

Color versions of one or more of the figures in this article are available online at http://ieeexplore.ieee.org

issue their own access control policies through contracts.

Access control is essentially the authorization and transfer of permissions, so the rights interaction records of resource visitors and resource authorizers (resource owners) are generated, and the access service records can be generated after recording these information. The access service record can not only trace the access of the resource and the permission interaction between the access subject and the object, but also analyze and judge a series of related assessments and malicious attacks through the record. There are many types of access control models (RBAC, ABAC, etc.). This paper constructs a distributed access service record book, named TokenChain, and evaluates the trust of resource visitors and resource authorizers through this data structure.

Trust management mechanism does not conflict with any kind of computer security mechanism, such as privacy protection, access control, congestion control, federation learning and so on. Gu et al. [4] mentioned a trust-based access control mechanism, which realizes the access hierarchical control policy by grading the trust level, and points out the possibility of combining trust management with access control. Wei et al. [7] gave a comprehensive overview of trust management in the Internet of things, discussed open problems and potential technologies, proposed a three-tier trust management architecture of data layer, computing layer and incentive layer, and pointed out that blockchain technology is the potential driving technology of trust management. Trust management mechanism has been applied to many aspects, including Internet of things trust management [2], and vehicle network (vehicle ad hoc network) trust management [8]. The existing trust value evaluation methods include probability methods (such as Bayesian network-based trust modeling, evidence theory, etc.), fuzzy logic [7] and so on. Kouicem et al. [2] described a distributed trust management protocol of the Internet of things, established a trust management model among Internet of things devices, service providers (SP) and fog nodes, weighted the last trust value, direct observation value and recommended value to calculate the trust evaluation, and tested the resistance performance under malicious attacks.

In the process of resource access and service authorization, the corresponding access service record information will be generated. Through the interactive information between these resources and services, we can carry out trust evaluation, and then achieve access control. The trust management scheme of the Internet of things based on access service records can realize multi-domain trust management through unified record data structures in a variety of scenarios, and can realize the quantitative control mechanism of access control. There are broad prospects for the management of a large number of resources in the Internet of things.

**Problem**

Wei [7] discusses in detail the following attribute requirements of existing trust management schemes for the Internet of things: security, effectiveness (time efficiency, etc.), reliability, attack resistance, flexibility, fairness, lightweight, distributed systems (such as blockchain), distributed, tamper-proof, consistency (consensus mechanism), etc., as well as timeliness, availability, etc. Wei points out that there are still the following problems with existing trust management schemes:

1) Bidirectional trust evaluation: At present, most trust management schemes are trust-evaluation methods of service requestor (SR) to a service provider (SP), but in practical application, SP also has corresponding trust evaluation to SR, and there is a lack of effective evaluation scheme for SR by SP. It is a challenge for trust management schemes to realize two-way trust evaluation between SR and SP at the same time.

2) Timely response to the dynamic change of the scene: in the vehicle accident information dissemination scenario, the dissemination of messages and the trust evaluation and detection of malicious devices need to be timely. Liu et al. [9] proposed a trust management scheme based on an anonymous vehicle announcement protocol, which evaluates the trust according to the data packet response of the vehicle where the accident occurred, and fails to identify the authenticity of the accident in time. Many existing trust management schemes fail to respond to the events in the scenario in time according to the dynamic changes in the scenario.

3) Privacy protection: the trust evaluation information in the trust management scheme should avoid malicious disclosure of information, and the privacy protection of users and devices based on service records and trust data is essential.

4) Cross-domain trust management: a variety of existing trust management schemes of the Internet of things are aimed at a single scenario, which has good results for specific scenarios, but can not be applied to other scenarios, such as [8] [9]. In multi-domain scenarios, authentication policies and trust management schemes can not achieve effective aggregation and synchronization, and the unified trust management scheme between different scenarios is a key research issue.

This paper proposes a trust management scheme based on access service records, which effectively solves the problems of two-way trust evaluation and cross-domain trust management, and its effectiveness is proved by simulation experiments. At the same time, the triple encryption mechanism based on DES [17] effectively protects the privacy of trust data, and the data of trust evaluation can be fed back to users, devices and services in time.

**Contribution**

The contribution of this paper consists of three parts, which are summarized as follows.

1) A multi-domain unified account book of distributed access service record data structure is proposed, which is called TokenChain. The service requester (SR), service provider (SP) and service (Service) are encrypted by a triple DES encryption algorithm [17], and two data sets are simulated, which are the fixed score data set and the malicious attack data set, and the movie score data set of two real scenes. The advantages and disadvantages of TokenChain and the performance of reading and writing are analyzed.

2) A trust management scheme based on TokenChain (TBTM) is proposed, which includes three layers of management architecture: data layer, computing layer and control layer. At the same time, a certain score of the trust value of the convergence analysis and related theoretical analysis (service satisfaction prediction, global trust level analysis), malicious attack resistance strategy and performance evaluation.

3) Through experiments on real data sets and other complex scenarios, the results show that the trust management scheme in this paper meets the requirements of multi-domain trust management. The trust management scheme in this paper is compared with other schemes, and its advantages are analyzed.

The organizational structure of this paper is as follows, the second section introduces the related work, and the third and fourth sections introduce the distributed access service record book (TokenChain) and the trust management scheme (TBTM) proposed in this paper. The fifth section carries on the theoretical analysis of this scheme, the sixth section analyzes the resistance to malicious attacks, and the seventh section evaluates the performance. Sections 8 and 9 analyze the feasibility of other complex scenarios and compare them with other trust management schemes. Finally, the work of this paper is summarized in section 10.

## II. RELATED WORK

### 2.1. Distributed Access Control

Centralized systems generally have problems such as poor scalability, prone to the single point of failure, and data easy to be tampered with. Distributed systems gradually replace the traditional centralized systems, and the Internet of things changes from traditional centralized servers to distributed servers. Access control has also changed from RBAC (role-based access control), ABAC (attribute-based access control) and CapBAC (power-based access control) to blockchain-based access control.

Novo et al. [1] proposed a lightweight scalable access control framework by combining the blockchain and the Internet of things, which carries out resource access control policies through the operations defined in the intelligent contract and the interfaces provided by the management nodes, including the definition, modification and development of policies. Resource owners can add access control rules by issuing contracts, and resource visitors can successfully access resources through intelligent contracts under predefined access conditions. Yuanyu et al. [14] proposed an access control framework based on intelligent contracts, which implements access control of Internet of things resources through registered contracts (RC), judgment contracts (JC) and multiple access control contracts (ACCs), and provides a complete case study, which shows the feasibility of distributed access control in the Internet of things.

In the process of distributed access control, the access subject applies to access the corresponding resources of the object, and the corresponding access service records will be generated after the successful access. In most access control frameworks, the data information in this area is not recorded, and the access to resources and the interaction history between the subject and object have not been fully recorded. At the same time, the credibility of access control between subject and object has not been quantitatively evaluated, and the entities of both sides of access authorization have not been quantitatively evaluated. this is a great challenge to the practical application and service management of the access control framework in the real scene of the Internet of things. Gu et al [4] mentioned a policy of access control grading based on entity trust level, which realizes the control of access resource application through the trust degree of the access subject, and then combines trust evaluation with access control.

Based on this problem, this paper proposes a distributed access service record data structure, an account book called TokenChain, which is used for trust management.

### 2.2. Distributed Trust Management in the Internet of things

Centralized trust management uses the traditional centralized server architecture to store, calculate and manage trust data information in the Internet of things. The general process is the interaction among the service requestor (SR), service provider (SP) and server. Centralized trust management scheme is gradually replaced by distributed trust management scheme because of the single point of failure. Wei et al. [7] pointed out that blockchain is a potential development technology of trust management, which makes a large number of untrusted entities form a trusted platform. Combined with the Hash function and Merkle tree to ensure the traceability and anti-tamper modification of historical service data. Distributed trust management of the Internet of things has a broad prospect in the application of trust data security and real scenarios of the Internet of things.

Kouicem et al. [2] proposed a distributed trust management protocol for the Internet of things based on blockchain, which evaluates the trust through the weighted calculation of the trust value, direct observation value and recommended value in the previous stage. The direct observation value is the mean of cumulative satisfaction, and the recommended value is calculated by fog nodes. Statistics include the average recommendation value provided by service providers and the weighted average recommendation value provided by the Internet of things devices. And the author analyzes the resistance strategies to cooperative malicious attacks, including malicious suggestion attacks and ballot-filling attacks. Among them, Kouicem applies a statistical analysis method of the satisfaction of the Internet of things equipment to the service provider, but does not form a specific evaluation standard of satisfaction, and only calculates its satisfaction as an observation. At the same time, there is no analysis of the feasibility and effectiveness of the application in the real-world networking scenario.

Yang et al. [8] proposed a distributed trust management scheme applied in vehicle networks. The probability of events is calculated by the Bayesian probability calculation method, and a positive score or a negative score is generated, and all scores are calculated to generate trust offset. Yang proposes a distributed consensus mechanism combining PoW and PoS based on trust offset, which uses the sum of absolute offsets as the equity value and sets an upper limit (to avoid being a bookkeeper if the equity value is too large). The difficulty of PoW (nonce) depends on the equity value to ensure that the data in the blockchain is updated in time. This consensus mechanism can be used in the incentive mechanism of distributed trust management. The higher the trust value, the lower the difficulty of mining, and the more likely it is to become a bookkeeper and promote equipment and service providers to provide better services.

Liu et al. [9] introduced an anonymous vehicle announcement protocol and a trust management model based on blockchain. Combined with the announcement protocol, RSU (roadside unit) is allowed to determine whether the message is true or not, and the trust value is calculated by logical regression. Liu implements the announcement protocol through the packet interaction among the initiator, responder and trusted authority (RSU), and ensures the anonymity and security of the data by using an anti-collision hash function and symmetric encryption algorithm. In the trust management model, two metrics, direct trust value (based on vehicle history behavior) and indirect trust value (based on recommendation, calculate the proportion of accident eyewitness vehicles with consent message in all witnesses) are used to evaluate, then the weighted calculation of reputation value is used to determine whether the message is credible, and then the consensus algorithm combined with PoW and PBFT is used to store its data in the blockchain. The scheme proposed by Liu can only be applied in the vehicle network, and when it is actually deployed on the highway, it is worth considering whether the vehicle performs the corresponding operation on the message, and the vehicle may have lazy behavior and refuse to implement it. The feasibility in the actual scenario of the Internet of things can not meet the actual needs.

This paper proposes a trust management model suitable for multi-domain and two-way trust management, and encourages service providers to provide better services. Importantly, through the data set of the real scene and the trust evaluation of multiple scenarios, it is proved that the scheme in this paper is feasible in real life, and meets the real needs of users and scenarios.

## III. DISTRIBUTED ACCESS SERVICE RECORD DATA STRUCTURE LEDGER: TOKENCHAIN

TokenChain is a distributed access service record book, and the data structure is similar to blockchain, and DES symmetric encryption algorithm [17] is used to protect data privacy, which meets the advantages of traceability, anonymity and Tamper proof. The symbolic terms and their descriptions that may be used in the context of this section are described in Table 1.

TABLE I

SYMBOLIC TERMS AND THEIR DESCRIPTIONS THAT MAY BE USED IN THE CONTEXT OF THIS SECTION

| | |
|---|---|
| $SR$、$SP$、$Service$ | Service Requestor, Service Provider, Service |
| $k_1$、$k_2$、$k_3$ | Three keys of the DES encryption algorithm |
| $E_{k_i}$、$D_{k_i}$ | DES encryption algorithm and decryption algorithm based on Secret key $k_i$ |
| $P(x)$、$C(x)$ | Plaintext and Ciphertext |
| $s$、$o$、$e$ | A service requester (access subject), service provider (access object), service (access transaction), $s \in SR$, $o \in SP$, $e \in Service$. |
| $S_{so}^e$ | s to the satisfaction score of the service e provided by o (different meanings can be given in different scenarios, such as time occupied, consumption, etc.), a real number. |
| $S_{max}$、$S_{min}$、$S_m$ | A maximum, minimum, and intermediate value of $S_{so}^e$ |
| $S_x$ | The objective satisfaction of an entity object $x$ ($x \in SR \cup SP \cup Service$), also known as trust offset. |

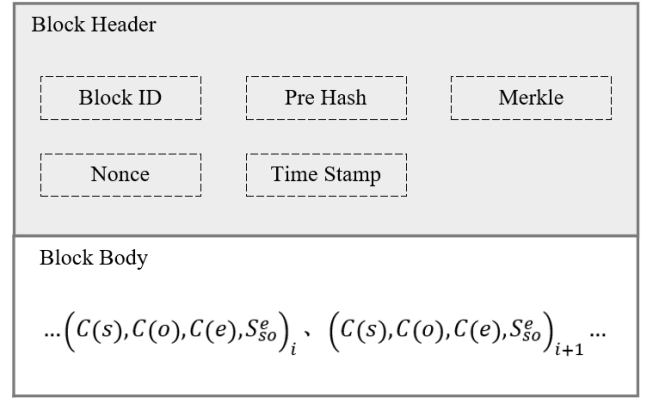

**Fig.1.** TokenChain data structure

### 3.1. Structure of TokenChain

By analyzing the data structure of the blockchain, the data structure of the distributed access service record is proposed. Its implementation account book TokenChain adopts the structure shown in figure 1. The block header is composed of ID, pre-hash, Merkle tree, nonce and time stamp (TimeStamp). The block is composed of access service records. The structure of each access service record is $(s, o, e, S_{so}^e)_i$, The ciphertext is encrypted by DES symmetric encryption algorithm, and its structure is $(C(s), C(o), C(e), S_{so}^e)_i$, and upload the encrypted data. The Geth client of the Ethernet Square platform is used to build a private chain platform, and PoW is used as a consensus mechanism to ensure data consistency. At the same time, TokenChain ensures the anonymity and privacy of data. In practical application scenarios, different practical meanings can be given to $s, o, e$ according to the actual situation. Used for real scenario assessment, such as customer's overall satisfaction with the merchant, can be set $e = NULL$. It can gives different practical meanings to $S_{so}^e$, such as the level of traffic accidents, the service time occupied by equipment and so on.

### 3.2. Data Protection

In order to protect the privacy, anonymity and security of data in TokenChain, the protection mechanism of data encryption is adopted in this paper. DES (Data Encryption Standard) encryption algorithm [17], which was promulgated and implemented by the American National Bureau of Standards in the 1970s, is a symmetric data encryption algorithm in the block cipher system. At present, the more effective attack method against the DES algorithm is exhaustive search, which has been widely used in many fields. Triple DES encryption is performed for $s, o, e$ of each access service record $(s, o, e, S_{so}^e)$, and a decryption method is given. The process of plaintext encryption to ciphertext and decryption is shown below, in which three different keys are used for encryption and decryption.

$$C = E_{k_3}(D_{k_2}(E_{k_1}(P))) \qquad (1)$$
$$P = D_{k_1}(E_{k_2}(D_{k_3}(C))) \qquad (2)$$

Eq. (1) is a data encryption and Eq. (2) is data decryption. The triple DES encryption algorithm based on three keys ensures the security of data and effectively prevents illegal attacks from divulging data privacy. At present, no one has found an effective attack method against this encryption scheme.

### 3.3. Data Set

In order to evaluate the trust management scheme more effectively, this paper prepares a total of four data sets, the first two data sets are simulation data sets, and the latter two data sets are movie rating data sets in real scenes [19-21]. The structures of the four data sets are all in line with the structure of TokenChain, and a triple DES encryption scheme is adopted. Note that the last two datasets are lost after encrypted storage, so this article uses anonymous ID fields for storage. Table 2 describes the details of the dataset. In the course of the experiment, due to the limitation of the computing power of the experimental equipment and the long waiting time in the process of running the program, only 8 million pieces of scoring data were run in dataset 3 and 4, and only 4 million pieces of scoring data were run in dataset 4.

TABLE II
DETAILS OF THE FOUR DATASETS

| Data Set | SR | SP | Service | $S_{so}^e$ | $S_{max}$ |
|---|---|---|---|---|---|
| dataset_1 | honest SR | honest SP | honest Service | 5 | 10 |
| dataset_2_1 | malicious SR | honest SP | honest Service | 1 | 10 |
| dataset_2_2 | malicious SR | malicious SP | malicious Service | 10 | 10 |
| dataset_3 | userID | tagID | movieID | - | 5 |
| dataset_4 | userID | tagID | movieID | - | 5 |

### 3.3.1. Dataset One

Honest service requesters (honestSR) give a fixed score to honest services (honestService) provided by honest service providers (honestSP). The score is $S_{so}^e = 5$ and $S_{max} = 10$, and the number of visits to the service record is $n$. Dataset 1 is a simulated honest access dataset, named dataset_1.

### 3.3.2. Dataset Two

The dataset is divided into two parts, namely, the malicious scoring attack dataset and the malicious ballot filling attack dataset, named dataset_2_1 and dataset_2_2. Their data simulations are as follows:

A. dataset_2_1: malicious service requesters (malicious SR) score a large number of malicious scores on services (honest Service) provided by honest service providers (honest SP). $S_{so}^e = 1$ and $S_{max} = 10$.

B. dataset_2_2: the malicious service requestor (malicious SR) populates a large number of ballots for the service (malicious Service) provided by the malicious service provider (malicious SP). $S_{so}^e = 10$ and $S_{max} = 10$.

The number of access service records of the two parts of the dataset is $n$, and the second dataset is the simulated malicious access attack dataset.

### 3.3.3. Dataset Three

Dataset 3 is the real scene movie rating dataset suitable for the new study, named dataset_3. The actual meaning of each access service record is (user ID, tag ID, movie ID, rating $S_{so}^e$). The tag ID is the label category to which the movie belongs. $S_{so}^e$ is the evaluation score given by the user to the corresponding movie. Among them, the minimum change range of $S_{so}^e$ is 0.5 and $S_{max} = 5$, a total of about 162000 users have scored 25000095 on 62000 films and 1128 tags, and the user's scores are truly collected.

### 3.3.4. Dataset Four

Dataset 4 is the movie rating dataset suitable for the education industry, named dataset_4, and the actual meaning of each access service record is equivalent to dataset 3. And the minimum change range of $S_{so}^e$ is 0.5 and $S_{max} = 5$. In dataset 4, a total of 280000 users scored 58000 movies with a score of 27753444, with a total of 1128 tags, and the dataset was collected in the real environment.

### 3.4. Performance Analysis

TokenChain is a distributed access service record book based on blockchain. The formal data structure ensures that TokenChain can be applied to multiple scenarios of the Internet of things, and inherits the advantages of the blockchain (only add, tamper-proof, traceability, data transparency), but also inherits the shortcomings of blockchain, such as false or private data can not be deleted and changed. TokenChain uses the DES encryption algorithm to achieve anonymity and protect the privacy of users. In terms of reading and writing performance, the adoption of the PoW consensus mechanism may have a high delay, and consensus mechanisms such as DBFT and PoS can be used to reduce latency [13].

## IV.     Trust Management Scheme：Tbtm

### 4.1. Overall Model Architecture

This paper proposes a trust management scheme based on TokenChain (TokenChain-Based Trust Management, TBTM), which can be applied in multi-domain scenarios of the Internet of things. TBTM consists of three layers, namely, the data layer, the computing layer and the control layer, and its specific architecture is shown in figure 2. The system reads the access service record data from TokenChain, registers the unregistered entity $(s, o, e)$, then evaluates the trust through the computing layer, stores the updated trust value information to the data layer, and the control layer controls some aspects of the overall scheme to ensure the effectiveness of TBTM. Table 3 summarizes the symbolic terms and their descriptions that may be used in this section.

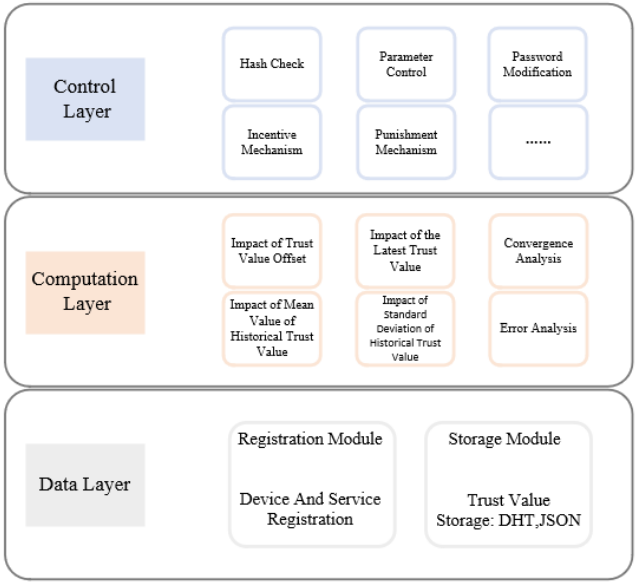

**Fig. 2.** TBTM overall model architecture

TABLE III
Symbolic Terms And Their Descriptions That May Be Used In This Section

| | |
|---|---|
| $PK$、$SK$ | The public key and private key of the registered entity |
| $T_i$、$T_0$、$Hash(T)$ | Trust value, initial trust value, historical trust value check value |
| $S_s$、$S_o$、$S_e$ | Trust offset of $s, o, e$ |
| $\bar{T}$、$D(T)$、$\sqrt{D(T)}$ | Mean, variance and standard deviation of historical trust values |
| $\alpha$、$\beta$、$\gamma$、$\delta$ | Weight parameter |
| $\lambda_s$、$\lambda_o$、$\lambda_e$ | Convergence error value of $s, o, e$ |
| $\kappa_s$、$\kappa_o$、$\kappa_e$ | The error of the average trust offset of $s, o, e$ |
| $\mu$、$\nu$、$\varepsilon$ | Threshold of early warning list, a threshold of the malicious list, early warning offset $(\mu > \nu)$ |
| $x$ | An entity $x \in (s, o, e)$ |

### 4.2. Data Layer

The data layer consists of two modules, namely the registration module and the storage model. The registration module is responsible for registering the devices and services, and the storage module is responsible for storing the information value data information into the distributed hash table (DHT) and the local JSON. At the same time, both adopt encrypted TokenChain data, and store private key and trust value check value through a one-way hash function. DHT is a distributed file storage system, which can be used for distributed cache and a large number of distributed file data storage. Without a central server, each client is responsible for storing a small part of the data, and its basic structure is key-value, which takes each trust data entry as value, its hash value as the key, and then stores this key-value pair to the distributed network, thus completing the storage and addressing of the whole DHT network. The DHT storage structure of the data is shown in figure 3.

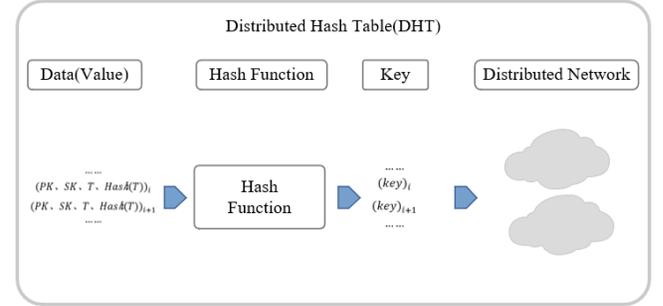

**Fig. 3.** DHT data structure

#### 4.2.1. Registration Module

The registration entity writes the public key and private key, and the initial trust value and historical trust value check value are generated by the system. The structure is $(PK$、$SK$、$T_0$、$Hash(T))$, which is stored in DHT. Entities include all users, devices, and services in TokenChain, namely SR, SP, and Service. When registering, each field is handled as follows:

1) PK: read all $s$、$o$、$e$ (encrypted data) in TokenChain to as a public key.
2) SK: user-defined and calculated by a one-way hash function to store its results, such as the SHA-256 hash function. Note that in order to save storage space in all the experiments in this article, "sk" is used instead of the private key for storage.
3) $T_0$: the system customizes the initial trust value, which will be described in detail below.
4) $Hash(T)$: the check value of the historical trust value, which has only the initial trust value of $T_0$ at the time of registration, so this value is $Hash(T_0)$.

#### 4.2.2. Storage Module

The storage of trust value data is divided into two parts, one is to use DHT to store trust value data according to $(PK$、$SK$、$T$、$Hash(T))$ format, the purpose is the distributed storage of data to ensure the security of data. The other part is to use JSON locally to store the format according to the key value of $(PK : [... T_i、 T_{i+1} ...])$, which is convenient for data statistics and analysis. at the same time, it checks the local data through $Hash(T)$, and reloads the local historical trust value data when it is found to have been tampered with. The calculation equation of $Hash(T)$ is shown below, and the method of the hash check is given in the control layer.

$$Hash(T_{i+1}) = Hash(Hash(T_i) + Hash(T_{i+1})) \qquad (3)$$

### 4.3. Computing Layer

The calculation layer includes three parts: trust evaluation, convergence analysis under certain conditions and error analysis. Trust evaluation is affected by four statistical indicators, and the trust value is obtained by weighted summation. Then in the convergence analysis, through a large number of simulation experiments of dataset 1 and dataset 2, the convergence results are obtained, and the influence and value of relevant parameters are determined. Finally, the error in the case of convergence is determined, and the error of the average trust offset is obtained.

#### 4.3.1. Trust Evaluation

The evaluation of the trust value is based on each access service record $(C(s), C(o), C(e), S_{so}^e)$ in the TokenChain. Note that the data is not decrypted during the trust value calculation. The calculation of trust value is affected by the following four statistics, and the updated trust value is obtained by weighted summation.

a. Trust offset

Read an access service record data $(C(s), C(o), C(e), S_{so}^e)$ in TokenChain. When the registration is completed, assign its $S_{so}^e$ to $s$、$o$、$e$ according to equation (4) (5). This can be changed according to the actual situation of the Internet of things scenario. $S_{so}^e$ pays more attention to the quality evaluation of $o$、$e$, and can encourage users to actively become service providers, promote the diversified development of services, and avoid the selfishness of network entities.

$$S_s : S_o : S_e = 1 : 3 : 2 \qquad (4)$$

$$S_s + S_o + S_e = \frac{S_{so}^e}{S_{max}} \in [0,1] \qquad (5)$$

$$S_s = (S_{so}^e / S_{max}) / 6 \qquad (6)$$

$$S_o = (S_{so}^e / S_{max}) / 2 \qquad (7)$$

$$S_e = (S_{so}^e / S_{max}) / 3 \qquad (8)$$

The trust offset of $s$、$o$、$e$ is calculated by equation (6) (7) (8). The weighted coefficient is $\alpha$ and the influence on trust evaluation is $\alpha S_x$. When $S_{so}^e = S_m$, $S_{min} = 0$ and $S_m = \frac{S_{max}+S_{min}}{2}$, $S_s = \frac{1}{12}$, $S_o = \frac{1}{4}$, $S_e = \frac{1}{6}$.

b. Latest trust value

The evaluation of trust value is affected by the latest trust value T, the weighted parameter is $\beta$, and its influence is $\beta T$.

c. Mean value of the historical trust

The evaluation of the trust value is affected by the historical trust value mean $\bar{T}$, the weighted parameter is $\gamma$, and its influence is $\gamma \bar{T}$. $\bar{T}$ is calculated by the following equation:

$$\bar{T} = \frac{1}{n} \sum_{i=1}^{n} T_i \qquad (9)$$

d. The standard deviation of historical trust value

The evaluation of trust value is affected by the fluctuation of historical trust value, which is affected by the overall standard deviation $\sqrt{D(T)}$, the weighted parameter is $\delta$, and the

influence is $\delta \sqrt{D(T)}$. The equation for calculating $\sqrt{D(T)}$ is as follows:

$$\sqrt{D(T)} = \sqrt{\frac{\sum_{i=1}^{n}(T_i - \bar{T})^2}{n}} = \sqrt{\frac{\sum_{i=1}^{n}(T_i)^2 - n\bar{T}^2}{n}} \qquad (10)$$

Therefore, the trust value of an entity ($s$、$o$、$e$) is calculated as follows:

$$T_{n+1} = \alpha S + \beta T_n + \gamma \bar{T} + \delta \sqrt{D(T)}$$
$$= \alpha S + \beta T_n + \gamma \frac{1}{n} \sum_{i=1}^{n} T_i + \delta \sqrt{\frac{\sum_{i=1}^{n}(T_i)^2 - n\bar{T}^2}{n}} \qquad (11)$$

And the four weight parameters $\alpha$、$\beta$、$\gamma$、$\delta$ meet the following requirements:

$$\begin{cases} \alpha > 0 \\ \beta > 0, \ \gamma > 0, \ \beta + \gamma = 1 \\ \delta < 0 \end{cases} \qquad (12)$$

#### 4.3.2. Convergence Analysis

According to the trust evaluation scheme, when $S_{so}^e$ is fixed, due to the influence of trust offset, the latest trust value, the mean value of historical trust value and the standard deviation of historical trust value, the trust value T will converge to a real number.

Lemma 1. Under the condition that $S_{so}^e$ is certain, when the number of access service records n tends to infinity, assuming that $T_n$ converges, then $\lim_{n\to\infty} T_{n+1} = \lim_{n\to\infty} T_n$.

Lemma 2. When $S_{so}^e$ is certain, when the number of access service records n tends to infinity, $\bar{T} = T_n$, that is, $\bar{T} = \lim_{n\to\infty} \frac{1}{n} \sum_{i=1}^{n} T_i = \lim_{n\to\infty} T_n$.

Theorem 1. When $S_{so}^e$ is certain, when the number of access service records n tends to infinity, $T_n$ converges to a specific real value $T_n'$, and the error is $\lambda$ ($T_n = T_n' + \lambda$).

$$T_n' = \sqrt{\frac{1}{n} \sum_{i=1}^{n} T_i^2 - \frac{\alpha^2 S^2}{\delta^2}} \qquad (13)$$

Proposition 1. When $S_{so}^e$ is certain, when the number of access service records n tends to infinity, the conditions that the trust value converges to the real number $T_n'$ are as follows: $\sum_{i=1}^{n} T_i^2 \geq \frac{n\alpha^2 S^2}{\delta^2}$

In order to verify the convergence of the parameters and determine the values of the parameters, a large number of simulations are carried out. The following eight groups of comparative experiments are prepared in this paper. each group statistics the trust value $T_n$ obtained by the trust evaluation equation (11) and the trust value $T_n'$ obtained by the convergence equation (13). The experimental results are shown in figure 4, the two trust values vary with the number of visits to the service n.

G1. $\alpha = 0.05, \beta = \gamma = 0.5, \delta = -0.5, T_0 = 0.1, dataset\_1$
G2. $\alpha = 0.01, \beta = \gamma = 0.5, \delta = -0.5, T_0 = 0.1, dataset\_1$
G3. $\alpha = 0.05, \beta = \gamma = 0.5, \delta = -1, T_0 = 0.1, dataset\_1$
G4. $\alpha = 0.05, \beta = 0.8, \gamma = 0.2, \delta = -0.5, T_0 = 0.1, dataset\_1$
G5. $\alpha = 0.05, \beta = 0.2, \gamma = 0.8, \delta = -0.5, T_0 = 0.1, dataset\_1$
G6. $\alpha = 0.05, \beta = \gamma = 0.5, \delta = -0.5, T_0 = 0.4, dataset\_1$
G7. $\alpha = 0.05, \beta = \gamma = 0.5, \delta = -0.5, T_0 = 0.1, dataset\_2\_1$
G8. $\alpha = 0.05, \beta = \gamma = 0.5, \delta = -0.5, T_0 = 0.1, dataset\_2\_2$

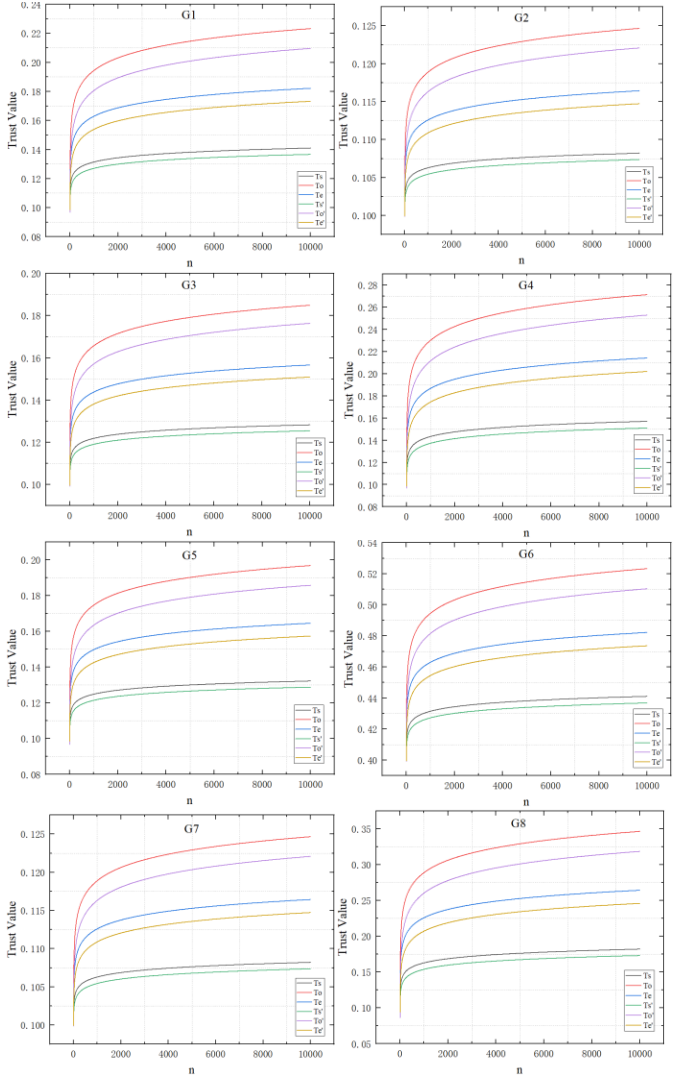

**Fig. 4.** Experimental results of G1-G8

Analysis of experimental results: through the analysis of experimental phenomena, the following results can be obtained:

a. G2 can be compared with G1: when $\alpha$ decreases, the overall trust value decreases.

b. G3 can be compared with G1: $\delta$ is smaller (the absolute value is larger) the overall trust value has decreased.

c. The comparative analysis of G4, G5 and G1 shows that the increase of trust value is larger at $\beta > \gamma$, and the larger the $\beta$ is, the greater the overall level of trust value is.

d. G6 can be compared with G1, and different $T_0$ has no effect on the trust value.

e. The comparative analysis of G7, G8 and G1 shows that the larger the $S_{so}^e$ is, the more the trust value increases.

f. By analyzing all the experimental results of G1-G8, it can be found that the trust values $T_n$ and $T_n'$ change greatly when $n < 1000$, and gently when $n > 1000$.

Therefore, through the comparative analysis of the above experiments, it is concluded that the value of the parameter should be $\alpha = 0.05, \beta = \gamma = 0.5, \delta = -0.5, T_0 = 0.1$. There are some limitations in the value of parameters, although a lot of experimental analysis has been carried out, but in the actual scene, the parameters may need to be changed because of the change of the real situation, so there is a parameter control module in the control layer of TBTM.

### 4.3.3. Error Analysis

In the above eight groups of comparative experiments, we can clearly see that there is a certain error between the trust evaluation value and the convergence value, and according to the experimental results, the error varies with $S_{so}^e$, and the error ratio of $s$、$o$、$e$ is about 1:3:2 under the same $S_{so}^e$. Therefore, the error of the experiments G1, G7 and G8 is analyzed, and the value of $\lambda_s$、$\lambda_o$、$\lambda_e$ and the ratio to $S_s$, $S_o$, $S_e$ respectively, that is, the value of $\kappa_s$、$\kappa_o$、$\kappa_e$, are calculated. The equation for $\lambda_x$、$\kappa_x$ is as follows, where $\kappa_x$ represents the error of the average trust offset. The experimental results of G1, G7 and G8 are shown in figure 5.

$$\lambda_x = (T_n - T_n') \tag{14}$$
$$\kappa_x = \lambda_x / S_x \tag{15}$$

Where $x \in (s、o、e)$.

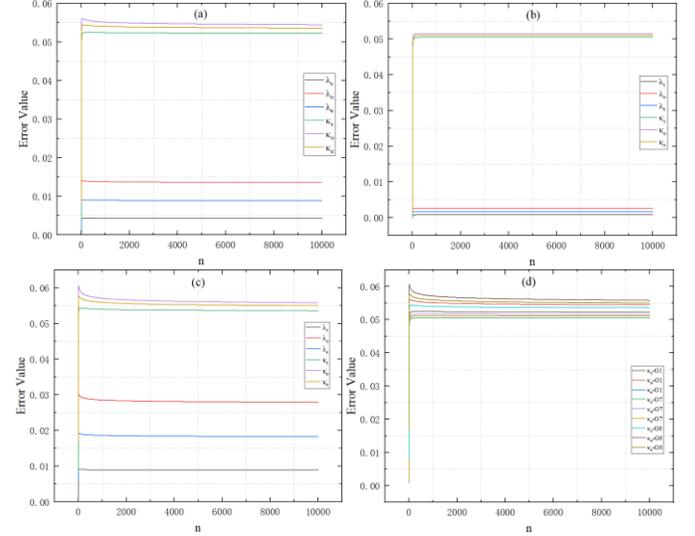

**Fig. 5.** results of experimental error analysis. (a) the experimental error analysis of G1, (b) the experimental error analysis of G7, and (c) the experimental error analysis of G8, all including $\lambda_s$、$\lambda_o$、$\lambda_e$ and $\kappa_s$、$\kappa_o$、$\kappa_e$. (d) is $\kappa_s$、$\kappa_o$、$\kappa_e$-analysis of the three. Notice that the curves of $\kappa_e - G1$ and $\kappa_s - G8$ are coincident.

Through the experimental results in figure 5, we can get that the average trust offset error $\kappa_x$ of $s$、$o$、$e$ in different $S_x$ of different data sets is almost the same, and the mean value of $\kappa$ is 0.05311685 through the experimental data. So $T_n = T_n' + \kappa S_x$.

### 4.4. Control Layer

The control layer of the TBTM model is shown in figure 2 and consists of five parts, each of which is described in detail below.

#### 4.4.1. Hash Check

The historical trust value is stored locally in the data layer, and the hash check value of the corresponding historical trust value is stored on the DHT. The calculation equation is (3). By calculating the local hash check value and comparing it with $Hash(T)$ in DHT, if different, it means that the local data has been modified and needs to be reloaded with the local historical trust value data. The hash check algorithm is shown below.

| Algorithm 1 | Hash Check |
|---|---|
| 1. | **for** $i$ from 0 to n |
| 2. | **if** i==0:$Hash(T)=Hash(T_0)$ |
| 3. | **else** $Hash(T)=Hash(Hash(T)+Hash(T_i))$ |
| 4. | **if** $Hash(T)$ is **false**: |
| 5. | **update** $Dict(PK)$ |
| 6. | **else return** True |

### 4.4.2. Parameter Control

The control layer can control the trust evaluation of TBTM through the dynamic control parameter $\alpha$、$\beta$、$\gamma$、$\delta$、$T_0$, which is helpful to adapt to different scenarios of the Internet of things.

### 4.4.3. Password Modification

An API for password modification is provided for users, through which users can modify their private key SK.

### 4.4.4. Incentive Mechanism

Two kinds of incentive mechanisms are adopted in TBTM, one is based on consensus agreement, and the other is based on the reverse auction. When the trust offset is allocated according to the score $S_{so}^e$, the allocation ratio of $S_s$、$S_o$、$S_e$ is 1:3:2. In different actual scenarios, the system can customize the allocation ratio to help motivate people to actively become SP and provide excellent service.

The purpose of the blockchain consensus mechanism [13] is to ensure the consistency of data in distributed systems. The most popular mechanisms are the PoW mechanism and PoS mechanism adopted by Bitcoin. The basic principle of PoW is that all the nodes participating in the consensus solve a common mathematical problem, and the first node to solve the problem broadcasts the calculation results. After being confirmed by the whole network, this node becomes the bookkeeper of this round, and is responsible for packing and linking the blocks. The bookkeeper's election relies heavily on the computing power of the node itself, and in most cases, the problem is to calculate a hash that meets the criteria (such as the previous bits are 0). The basic principle of PoS is that the bookkeeper is elected depending on the equity value of the node, which saves the calculation cost, but easily leads to centralization (the bookkeeping right is controlled by the node with a larger equity value).

In the incentive mechanism based on the consensus agreement, this paper adopts the consensus agreement of the combination of PoW and PoS proposed by Yang [8]. The trust value of the entity is regarded as the equity value, and its equity value determines the mining difficulty (nonce). The higher the equity value, the lower the mining difficulty, and the more likely it is to become a bookkeeper. Set an initial difficulty $nonce_0$ for all consensus nodes, and the $nonce$ value decreases with the increase of T, that is, the higher the trust value, the lower the difficulty of mining, and the entities are encouraged to provide better services to obtain higher trust values.

The incentive mechanism of home page recommendation based on reverse auction mainly applies the principles of reverse auction and service satisfaction prediction. The higher the trust value is, the more likely the service provider is to get the recommendation. In order to help service providers to achieve greater economic benefits in the actual scenario, Zhu [22] proposes a fair incentive mechanism in group perception based on reverse auctions and Vickrey auctions. The requestor sends a service request to the platform, and the platform selects the provider through the incentive mechanism and group bidding, and carries on the payment decision and quality review. Reverse auction means that buyers provide product information, demand for services and affordable price positioning, and sellers determine the final product providers and service providers in a competitive way. so that the buyer can realize the purchase with the best performance-price ratio.

The principles and methods of service satisfaction prediction are introduced in detail in Section 5. the trust value of $s$、$o$、$e$ is used to predict the $s$'s prediction score $P_{so}^e$ of the service $e$ provided by $o$, so as to determine the recommendation of $o$、$e$. The incentive mechanism based on reverse auction is mainly to give a predicted value of service satisfaction through $s$, and when the predicted value $P_{so}^e$ of the corresponding $o$、$e$ obtained by the system meets the requirements, the system recommends $o$、$e$ in order to motivate users.

### 4.4.5. Punishment Mechanism

The punishment mechanism of the TBTM model is mainly to effectively manage malicious devices or malicious services to enhance the security of the system. The system sets up an early warning list and a malicious list, and sets the corresponding early warning threshold $\mu$ and malicious threshold $\nu(\mu > \nu)$. Note that $\mu$、$\nu$ varies dynamically according to the global trust level. When the trust value of an entity is lower than $\mu$, the system adds it to the early warning list, and the entities in the early warning list add an early warning offset $\varepsilon(\varepsilon < 0)$ at the same time the system issues a warning. When the trust value is lower than $\nu$, add it to the malicious list and prohibit all services related to this entity. The specific criteria are as follows:

$$\begin{cases} T_n > \mu: \ Normal \\ \nu < T_n \leq \mu: \ Early \ warning \ list, add \ \varepsilon \\ T_n \leq \nu: \ Malicious \ list \end{cases}$$

## V. THEORETICAL ANALYSIS

### 5.1. Prediction of Service Satisfaction

The service satisfaction value $P_{so}^e$ is estimated by the historical trust value of $s$、$o$、$e$, and compared with $S_{so}^e$, this paper predicts and analyzes the service satisfaction of datasets 1, 2, 3 and 4 respectively, and measures its error.

Hypothesis 1. Suppose $P_{so}^e \propto T_s$、$T_o$、$T_e$ is true, get $P_{so}^e = aT_s + bT_o + cT_e + \omega$.

When $S_{so}^e$ is constant, the trust value of n access service records is evaluated, $T_s$、$T_o$、$T_e$ is dynamic, so there must be two times in which the value of $a$、$b$、$c$、$\omega$ is different, so hypothesis 1 is not true.

Theorem 2. When $S_{so}^e$ is certain, when the number of access service records n tends to infinity, the predicted value of service satisfaction is:

$$P_{so}^e = -\frac{\delta}{\alpha} S_{max} \left( \sqrt{\frac{1}{n}\sum_{i=1}^n T_i^2 - (T_n - \lambda)^2} \right)_s +$$

$$-\frac{\delta}{\alpha} S_{max} \left( \sqrt{\frac{1}{n}\sum_{i=1}^n T_i^2 - (T_n - \lambda)^2} \right)_o +$$

$$-\frac{\delta}{\alpha} S_{max} \left( \sqrt{\frac{1}{n}\sum_{i=1}^n T_i^2 - (T_n - \lambda)^2} \right)_e$$

$$\omega = S_{so}^e - P_{so}^e \qquad (16)$$

Where $\lambda_x = \kappa S_x$, $\omega$ is the error value.

Dataset 1 and dataset 2 are $S_{so}^e$-fixed data sets, so the size of the error value $\omega$ obtained by Theorem 2 in these two datasets, and the changes of service forecast satisfaction $P_{so}^e$ and true score $S_{so}^e$, the results are shown in figure 6. At the same time, $S_{so}^e$ in dataset 3 and 4 is a dynamically changing real dataset. This paper analyzes the change of the error value $\omega$ obtained by Theorem 2 in datasets 3 and 4. Because $S_{so}^e$ and $P_{so}^e$ are discrete and the true score of $S_{so}^e$ has no change rule, it is not convenient to show the corresponding change relationship between $S_{so}^e$ and $P_{so}^e$. Note that $S_{max} = 5$ in dataset 3 and 4, sometimes the value of $P_{so}^e$ is NaN (not a number) in datasets 3 and 4, which is not counted. The experimental results in figure 7 reflect that the error of the predicted value is about -2 to 2, indicating that the data trust evaluation in the same practical sense in the field of education and new research can be intercommunicated.

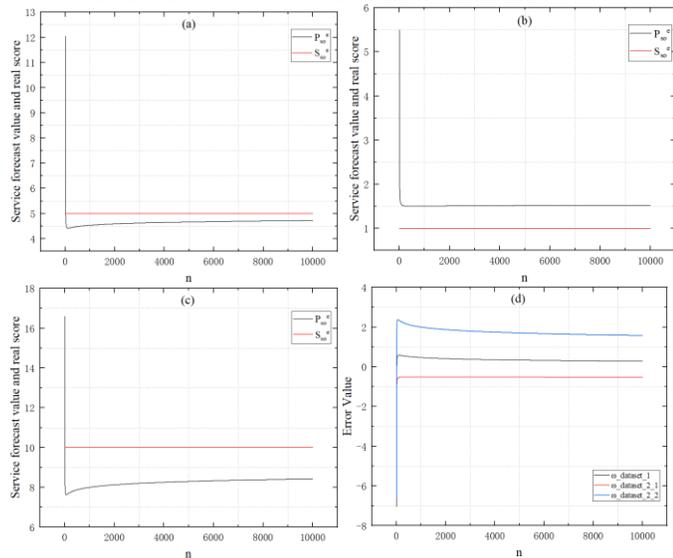

**Fig. 6.** The experimental results of Theorem 2 in Dataset 1 and Dataset 2: (a), (b) and (c) are the dynamic changes of $S_{so}^e$ and $P_{so}^e$ with the number of accesses, and (a) is dataset_1, (b) is a dataset_2_1, (c) is a dataset_2_2. (d) It is the error analysis result of three experiments.

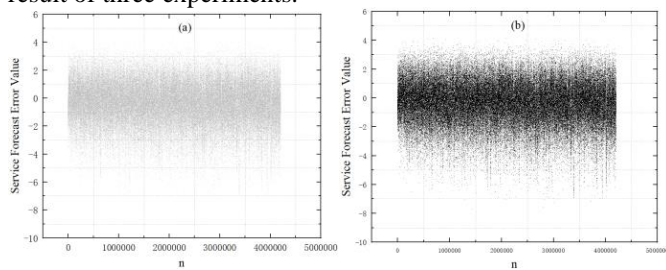

**Fig. 7.** Theorem 2 Experimental results (error values) in datasets 3 (a) and 4 (b)

### 5.2. Analysis of Global Trust Level

In the two-way trust evaluation, the distribution of the latest trust value of each entity of SR, SP and Service is analyzed. Figure 8 reflects the trust value distribution of datasets 3 and 4 after the trust evaluation of the TBTM model. The Abscissa is the entity label and the ordinate is the trust value of the entity. In TokenChain datasets 3 and 4, only part of the data completed trust evaluation, but all entities were counted, so there are a large number of entities whose trust values are initialized in the experimental results, the value is 0.1. SR, SP and Service refer to userID, tagID and movieID respectively. The experimental results in figure 8 show that the same attribute data in different domains can be evaluated by TBTM.

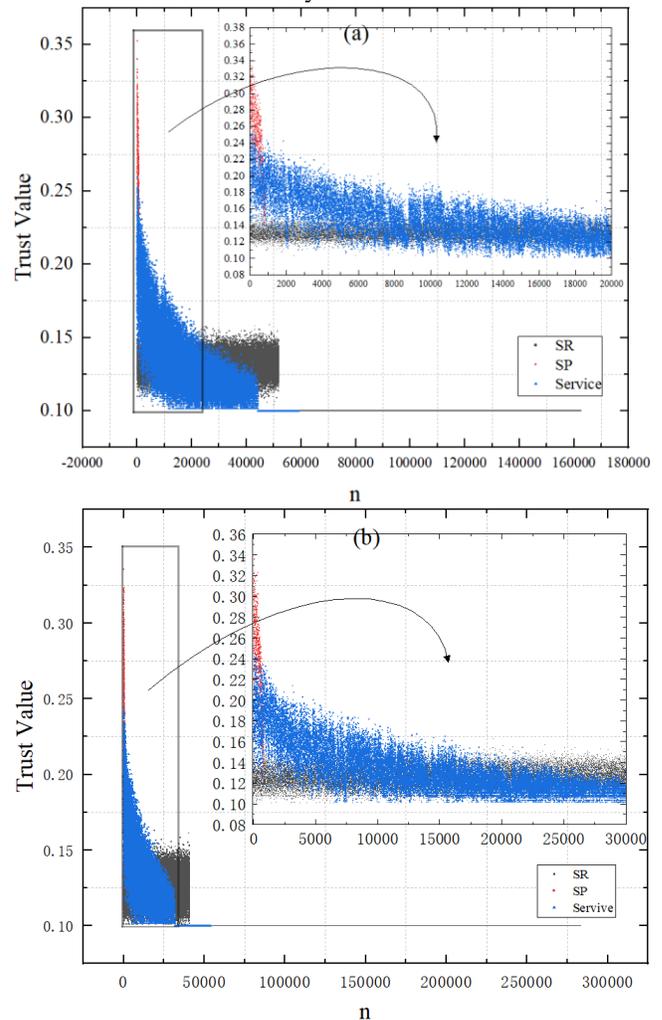

**Fig. 8.** The global distribution of trust levels, (a) is dataset 3, and (b) is dataset 4.

### VI. MALICIOUS ATTACK RESISTANCE STRATEGY

In the real physical networking scene, for the security of the system, it is necessary to avoid a variety of network attacks, protect data privacy and prevent data leakage, so that the attack of illegals can not succeed. There are many kinds of malicious requests and malicious attacks. In this paper, the following situations are discussed and analyzed in detail and the corresponding malicious attack resistance strategies are developed.

## 6.1. Selfish Node

Selfish node means that selfish users or selfish devices only request services but do not provide services (in the scenario where SR can become SP, the four data sets given in this paper are not compliant), which reduces the overall quality of service in the network and has a negative effect on the diversified development of service vendors and types of services. If there is no incentive mechanism in a system, it is likely to lead to "hitchhiking behavior" [23], which means that a node in a P2P network only uses network services and does not contribute to the network, so the incentive mechanism of the system is very necessary. The incentive mechanism in this paper meets the requirements of the system very well.

According to equation (4), $S_o = 3S_s$ and $S_e = 2S_s$, the trust offset of the service provider is the largest, the trust offset of the service provider is the second, and the trust offset of the service requester is the smallest, which helps to motivate users to become SP actively. Figure 4 also shows that $T_o > T_e > T_s$ under the same condition, and the trust difference is large. Therefore, when the device does not provide services, its trust value will be in a low state and will not get the accounting rights in the consensus mechanism and the opportunity for home page recommendation. The situation of selfish nodes is mainly used in the Internet of things scenario of two-way trust evaluation, and the TBTM model scheme meets the requirements of two-way trust evaluation and the prevention and control strategy of node selfishness.

## 6.2. Replay Attack

Attackers can repeatedly generate access service records, which interferes with the reliability of the trust assessment and increases the useless overhead of the network. Liu [9] effectively prevents replay attacks through a unique time parameter, so a minimum period parameter $\tau$ can be set in the TBTM model. When there are two or more identical access service records in $\tau$, it can be judged as a replay attack. The system only calculates exactly the same access service record data once to effectively resist replay attacks.

## 6.3. Malicious Scoring Attack (Bad-Mouthing Attacks)

A malicious scoring attack is a malicious SR that gives a very low score to the Service provided by an honest SP (e. G. peer review), and dataset_2_1 is a malicious scoring dataset. When there is one or a small number of very low scores, it is difficult to determine whether it is a malicious attack or a normal one, but when multiple SRs perform a large number of very low scores on a particular SP (or even a particular service), it must be judged to be a malicious scoring attack, so it usually occurs as a collaborative attack. In the TBTM model, the same score will affect not only the trust value of SP and Service, but also the trust value of SR, so it has a certain containment effect on the very low score. And through the G7 in figure 4, we can see that TBTM has a certain resistance to a large number of malicious scoring attacks, and the trust value will eventually converge to a real number.

## 6.4. Ballot Filling Attack

A ballot-filling attack is a malicious SR that gives a very high score (ballot filling) to the Service provided by a malicious SP. Dataset_2_2 is a ballot-filling attack dataset. Like malicious scoring attacks, a small number of attacks cannot be judged, and a large number of cooperative ballot-filling attacks are generally used to judge. As can be seen from G8 in figure 4, TBTM has some control over a large number of filled ballots, and the trust value will eventually converge to a real value. At the same time, the system can detect the massive filling of votes through the replay attack detection mechanism, and then shield the redundant scores (only count the same data once).

## 6.5. Cooperative Attack

Cooperative attacks are usually in the form of multi-party combination attacks, in which a group of nodes carries out malicious scoring attacks or ballot-filling attacks on a specific node (or even a specific service), similar to brushing. This situation may be classified into two types, one set of nodes is malicious or this particular node is malicious. TBTM is very resistant to cooperative attacks. As shown in G7 and G8 in figure 4, the change of trust value will become smaller and smaller with the continuous growth of access service records.

## 6.6. On-off Attacks

The on-off attack means that a node randomly provides good service or bad service, and in TokenChain, it shows that a user scores randomly, which is a serious threat to trust management. In this paper, the following four kinds of On-off attacks are simulated by Theorem 3, and the change of $s$、$o$、$e$ trust value is analyzed, the change period is $C$, open means extremely high score (experimental score is 10), off means extremely low score (experimental score is 1), and $S_{max} = 10$. The experimental results are shown in figure 9, and the value of $C$ in the experiment is 200 access service records. Through the experimental results, we can see that the TBTM model has a certain resistance to On-off attacks, and the trust value is always in a certain range.

The main form of On-off attack is to obtain high trust value through excellent service over a period of time, and then providing bad service on the basis of high trust value is still accepted by the group. The experimental results of figure 9 show that the resistance of the TBTM model to the On-off attack is timely, and the higher the switching frequency, the more timely the trust feedback of the system, which effectively resists the timely response to the bad service based on the high trust value under the On-off attack.

Theorem 3: there is a real number $a$、$b$ such that

$$(1+a)\%b = 10 \tag{17}$$

$$(10+a)\%b = 1 \tag{18}$$

If it is true, $b = 18, a = 18m + 9 (m \in N)$

    a.   Turn it on 100 times and then turn it off 100 times in $C$.

    b.   Turn it off 100 times and then turn it on 100 times in $C$.

    c.   50 switch combinations in $C$.

    d.   The switch appears randomly in $C$.

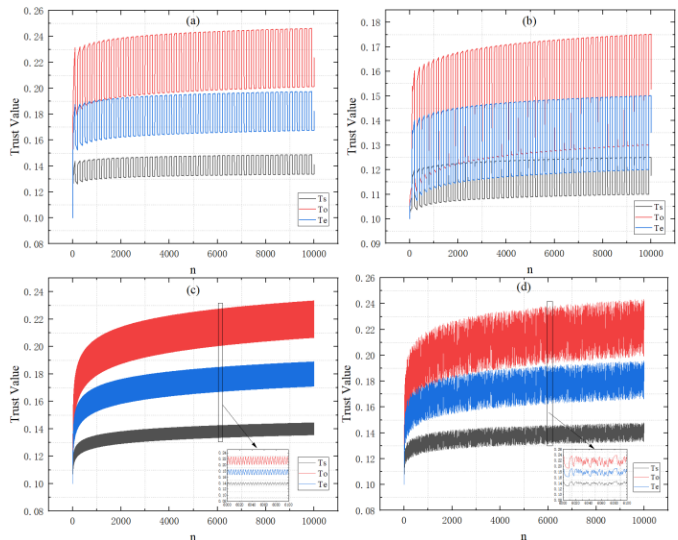

**Fig. 9.** The change of trust value of TBTM under On-off attack

### 6.7. Message Fraud Attack

Message fraud attack means that malicious devices deliberately generate false services and scores. Due to the characteristics of TokenChain, once the data is uploaded, it cannot be changed and deleted. For false data, corresponding defense measures can only be taken in the data layer and computing layer of the TBTM model to confirm the transaction through the digital signature of the user's private key (including entity registration and entity trust value update). Or analyze and check before TokenChain uploads to chains, such as judging the trust value of $s$、$o$、$e$, or the relationship between $s$、$o$、$e$, or history record, and so on.

### 6.8. Trust Value Tampering Attack

TokenChain uses a similar blockchain structure for storage, and trust value data information is stored by DHT. Both of them ensure data traceability and are tamper-proof through Merkle and hash functions. The local historical trust data is checked by the hash verification mechanism of the control layer to effectively avoid data tampering attacks.

## VII. PERFORMANCE EVALUATION

This paper analyzes the time efficiency of the TBTM model and mainly calculates the time consumption of trust evaluation in the calculation layer, that is, the time consumption of trust value calculation. The main process of trust value update is to read the file from TokenChain, then calculate the trust value (unregistered entities register), and finally store the trust-related data in the corresponding file. The calculation of the trust value takes a lot of time, and figure 10 reflects the time consumed by the calculation of the trust value (the first 1 million pieces of data). Note that the experimental environment of all the experiments in this paper is 11thGenIntel (R) Core (TM) i7-1165G7@2.80GHz, the memory is 16G, the graphics card is MX450, the programming language used in the experiment is python, and GPU is not applied to the experiment, only CPU is used to process the python program.

From the experimental results, there is a linear relationship between the time consumption of trust evaluation and the number of data bars. In practical application, the amount of historical trust value data of local JSON files can be limited on the basis of not affecting the overall trust evaluation as much as possible, so as to reduce the time consumption.

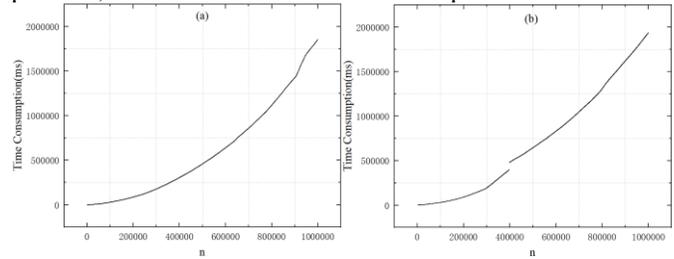

**Fig. 10.** Trust value calculation time consumption, (a) and (b) are the experimental results of data 3 and 4, respectively.

## VIII. EXPERIMENTS IN OTHER COMPLEX SCENARIOS

In order to verify the effectiveness of the trust management scheme proposed in this paper in different domains and the feasibility of adapting to complex scenarios, a small data set of sensor usage time is collected in the surrounding area, which is named dataset_5.dat. where $S_{max} = 24$, The time range of collecting data is 24 hours, and the data structure of collected data is (using device/user, sensor type, sensor, using time). There are 12 pieces of data in this dataset, as shown in figure 11, of which there are 26 entities.

The experimental results are shown in figure 12, and figure 12.a. shows the comparison between the time prediction and the real value. We can see that there is roughly the same trend between them. From figure 6, we can see that the more experimental data, the more accurate the results. Figure 12.b. shows the results of the trust evaluation, from which you can see that the trust value of LightClass is the highest, which means that LightClass has the highest usage time (frequency). Through the simulation experiment of a sensor using time data set, it is shown that TBTM is still effective in other complex Internet of things scenarios.

| 1. | Energy-Saving Lamp | Noise Class | Sound Sensor | 18 |
|---|---|---|---|---|
| 2. | Energy-Saving Lamp | Light Class | Light Sensor | 24 |
| 3. | Crowd | Space Class | Distance Sensor | 18 |
| 4. | Mobile Devic | Strength Class | Gravity Sensor | 24 |
| 5. | Mobile Devic | Space Class | GPS | 24 |
| 6. | Mobile Devic | Light Class | Light Sensor | 24 |
| 7. | Fan | Temperature Class | Temperature Sensor | 24 |
| 8. | Null | Humidity Class | Humidity Sensor | 24 |
| 9. | Camera | Light Class | Infrared Sensor | 24 |
| 10. | Touchable Device | Strength Class | Pressure Sensor | 18 |
| 11. | Crowd | Image Class | Image Sensor | 18 |
| 12. | Camera | Color Class | Color Sensor | 24 |

**Fig. 11.** Contents of dataset_5.dat

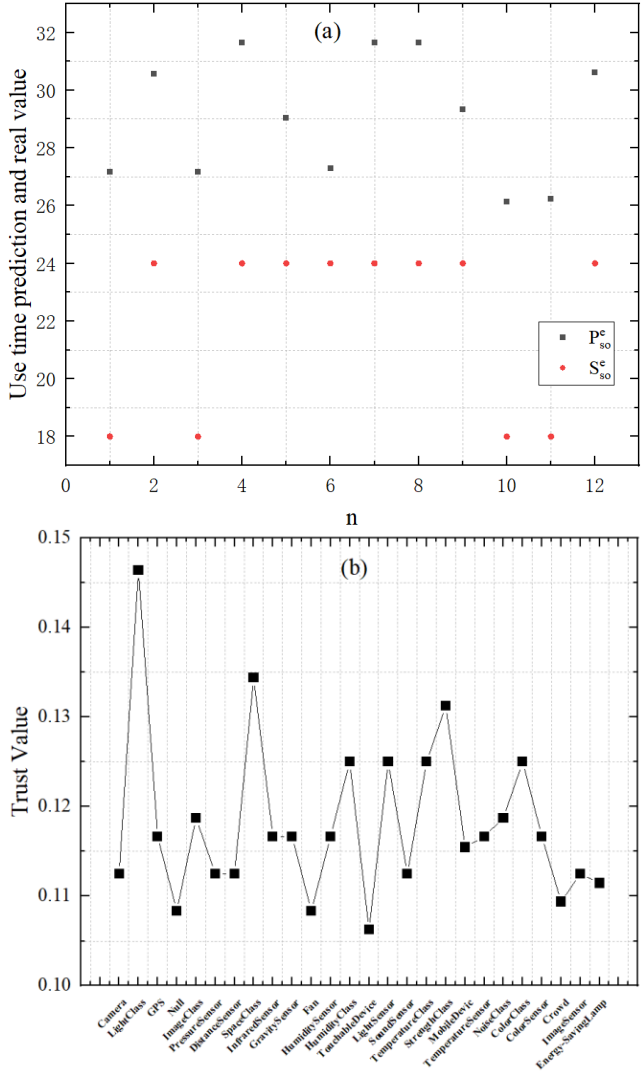

**Fig. 12.** Experimental analysis results

## IX. Comparison With Other Trust Management Schemes

This article mainly compares the different aspects of other trust management schemes, and the results are shown in Table 4, and Wei [7] describes more. Because TBTM calculates data and calculates trust values based on TokenChain, and stores trust-related data in the data layer, including DHT and JSON, TBTM does not meet the lightweight characteristics. However, the storage burden can be reduced through the distributed storage of data and the network composed of a large number of distributed nodes.

### TABLE IV
### Comparison Between TBTM And Other Trust Management Schemes

| Attribute | TBTM | [2] | [8] | [9] |
|---|---|---|---|---|
| Security | ✓ | ✓ | ✓ | ✓ |
| Validity | ✓ | | ✓ | ✓ |
| Reliability | ✓ | | ✓ | ✓ |
| Attack resistance | ✓ | ✓ | ✓ | ✓ |
| Flexibility | ✓ | | | |
| Fairness | ✓ | ✓ | ✓ | ✓ |
| Lightweight | | ✓ | ✓ | |
| Distributed | ✓ | ✓ | ✓ | ✓ |
| Tamper-proof | ✓ | ✓ | ✓ | ✓ |
| Consistency | ✓ | ✓ | ✓ | ✓ |
| Timeliness | ✓ | ✓ | ✓ | ✓ |
| Two-way evaluation | ✓ | | | |
| Privacy protection | ✓ | | ✓ | ✓ |
| Cross-domain | ✓ | | | |

## X. Conclusion

This paper proposes a TokenChain data structure and a trust management scheme (TBTM) based on this structure. TBTM can manage the trust of different domains according to the diversified attributes of TokenChain, and evaluate the trust of entities in each domain. Trust management mechanisms can be used as an auxiliary scheme for other mechanisms. At the same time, this paper gives the prediction scheme of attribute value $S_{so}^e$ and the global trust value distribution. Simulation experiments in other complex scenarios determine the effectiveness of TBTM in complex multi-domain IoT scenarios. Performance analysis determines that the time consumed is linearly proportional to the number of data, and determines the availability of TBTM in large-scale IoT scenarios. TokenChain and TBTM can be given different practical meanings according to different attributes in the real-world networking scene, which is still valid.

## Appendix

### A. Proof of Lemma 2

According to Lemma 1, when n tends to infinity.

$$\lim_{n \to \infty} T_{n+m+1} = \lim_{n \to \infty} T_{n+m} \ (m \in N)$$

$$\bar{T} = \lim_{n \to \infty} \frac{1}{m+1} \sum_{i=n}^{n+m} T_i$$

$$= \lim_{n \to \infty} \frac{1}{m+1} (m+1) T_n$$

$$= \lim_{n \to \infty} T_n$$

Therefore:

$$\bar{T} = \lim_{n \to \infty} \frac{1}{n} \sum_{i=1}^{n} T_i = \lim_{n \to \infty} T_n$$

### B. Proof of Theorem 1

When n tends to infinity, according to Lemma 2, get:

$$T_{n+1} = \alpha S + \beta T_n + \gamma \bar{T} + \delta \sqrt{D(T)}$$

$$= \alpha S + \beta T_n + \gamma \frac{1}{n} \sum_{i=1}^{n} T_i + \delta \sqrt{\frac{\sum_{i=1}^{n}(T_i)^2 - n\bar{T}^2}{n}}$$

$$= \alpha S + \beta T_n + \gamma T_n + \delta \sqrt{\frac{\sum_{i=1}^{n}(T_i)^2 - nT_n^2}{n}}$$

Since $\beta + \gamma = 1$ and Lemma 1: $\lim_{n \to \infty} T_{n+1} = \lim_{n \to \infty} T_n$ are obtained:

$$\alpha S + \delta \sqrt{\frac{\sum_{i=1}^{n}(T_i)^2 - nT_n^2}{n}} = 0$$

$$\alpha^2 S^2 = \delta^2 \frac{\sum_{i=1}^{n}(T_i)^2 - nT_n^{\ 2}}{n}$$

$$n\alpha^2 S^2 = \delta^2 \left(\sum_{i=1}^{n}(T_i)^2 - nT_n^{\ 2}\right)$$

$$\frac{n\alpha^2 S^2}{\delta^2} = \sum_{i=1}^{n}(T_i)^2 - nT_n^{\ 2}$$

Get:

$$T_n = \sqrt{\frac{\sum_{i=1}^{n}(T_i)^2 - \frac{n\alpha^2 S^2}{\delta^2}}{n}}$$

$$= \sqrt{\frac{1}{n}\sum_{i=1}^{n}(T_i)^2 - \frac{\alpha^2 S^2}{\delta^2}}$$

Therefore:

$$T_n' = \sqrt{\frac{1}{n}\sum_{i=1}^{n}T_i^2 - \frac{\alpha^2 S^2}{\delta^2}}$$

Because Lemma 1 and Lemma 2 are not valid when the trust value evaluation is in the initial stage, there is a convergence error $\lambda = T_n - T_n'$.

C.  Argumentation of Proposition 1

When the trust value converges to a real number, $T_n'$ is a real number, so

$$\frac{1}{n}\sum_{i=1}^{n}T_i^2 - \frac{\alpha^2 S^2}{\delta^2} \geq 0$$

Therefore:

$$\sum_{i=1}^{n}T_i^2 \geq \frac{n\alpha^2 S^2}{\delta^2}$$

Where $S \in [0,0.5]$, the experimental G1-G8 all meet this requirement.

D.  Proof of Theorem 2

According to Theorem 1, get:

$$T_n = T_n' + \lambda = \sqrt{\frac{1}{n}\sum_{i=1}^{n}T_i^2 - \frac{\alpha^2 S^2}{\delta^2}} + \lambda$$

The reasoning is:

$$\frac{1}{n}\sum_{i=1}^{n}T_i^2 - \frac{\alpha^2 S^2}{\delta^2} = (T_n - \lambda)^2$$

$$\frac{\alpha^2 S^2}{\delta^2} = \frac{1}{n}\sum_{i=1}^{n}T_i^2 - (T_n - \lambda)^2$$

$$S^2 = \frac{\delta^2}{\alpha^2}\left(\frac{1}{n}\sum_{i=1}^{n}T_i^2 - (T_n - \lambda)^2\right)$$

$$S = \frac{-\delta}{\alpha}\sqrt{\frac{1}{n}\sum_{i=1}^{n}T_i^2 - (T_n - \lambda)^2}$$

And because of equation (5):

$$S_{so}^e = (S_s + S_o + S_e)S_{max}$$

So, get:

$$S_{so}^e = \frac{-\delta}{\alpha}S_{max}\left(\sqrt{\frac{1}{n}\sum_{i=1}^{n}T_i^2 - (T_n - \lambda)^2}\right)_s +$$

$$\frac{-\delta}{\alpha}S_{max}\left(\sqrt{\frac{1}{n}\sum_{i=1}^{n}T_i^2 - (T_n - \lambda)^2}\right)_o +$$

$$\frac{-\delta}{\alpha}S_{max}\left(\sqrt{\frac{1}{n}\sum_{i=1}^{n}T_i^2 - (T_n - \lambda)^2}\right)_e + \omega$$

Where $\omega$ is the error value, so:

$$S_{so}^e = P_{so}^e + \omega$$

According to the error analysis of the experimental results in figure 5, the following results are obtained:

$$T_n = T_n' + \lambda_x = T_n' + \kappa S_x$$

That is:

$$\lambda_x = \kappa S_x$$

As a result, get:

$$P_{so}^e = \frac{-\delta}{\alpha}S_{max}\left(\sqrt{\frac{1}{n}\sum_{i=1}^{n}T_i^2 - (T_n - \kappa S)^2}\right)_s +$$

$$\frac{-\delta}{\alpha}S_{max}\left(\sqrt{\frac{1}{n}\sum_{i=1}^{n}T_i^2 - (T_n - \kappa S)^2}\right)_o +$$

$$\frac{-\delta}{\alpha}S_{max}\left(\sqrt{\frac{1}{n}\sum_{i=1}^{n}T_i^2 - (T_n - \kappa S)^2}\right)_e$$

E.  Proof of Theorem 3

The existence of a real number $a, b > 0$ makes the following two equations true.

$$(1 + a)\%b = 10$$
$$(10 + a)\%b = 1$$

Get $b > 10$, let $m, n \in N$, then

$$(1 + a) = mb + 10$$
$$(10 + a) = nb + 1$$

Then:

$$a = mb + 9 = nb - 9$$
$$(n - m)b = 18$$

$b$ is a factor of 18, then $b = 18$.
So $a = 18m + 9 (m \in N)$

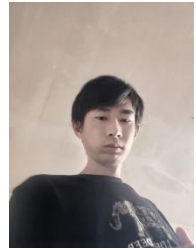

**Xuefei Li,** born in April 1999, received his bachelor's degree from Inner Mongolia University in 2021 and is currently studying at Inner Mongolia University in Hohhot, China.

His research interests mainly include distributed Internet of things, access control, trust management and so on.

**Ru Li**, born in 1974, Ph.D., is currently a professor at Inner Mongolia University in Hohhot, China. Her research interests mainly include the next-generation Internet and so on.